\documentclass[12pt]{iopart}
\usepackage{iopams}
\begin{document}

\title[Entanglement detection and quantification from the correlation matrix criterion]{Further results on entanglement detection and quantification from the correlation matrix criterion}

\author{Julio I de Vicente}

\address{Departamento de Matem\'aticas, Universidad Carlos III
de Madrid, Avda.\ de la Universidad 30, 28911 Legan\'es, Madrid,
Spain} \ead{jdvicent@math.uc3m.es}
\begin{abstract}
The correlation matrix (CM) criterion is a recently derived powerful
sufficient condition for the presence of entanglement in bipartite
quantum states of arbitrary dimensions. It has been shown that it
can be stronger than the positive partial transpose (PPT) criterion,
as well as the computable cross norm or realignment (CCNR) criterion
in different situations. However, it remained as an open question
whether there existed sets of states for which the CM criterion
could be stronger than both criteria simultaneously. Here, we give
an affirmative answer to this question by providing examples of
entangled states that scape detection by both the PPT and CCNR
criteria whose entanglement is revealed by the CM condition. We also
show that the CM can be used to measure the entanglement of pure
states and obtain lower bounds for the entanglement measure known as
tangle for general (mixed) states.
\end{abstract}

\submitto{\JPA}
\maketitle

\section{Introduction}

Entanglement in composite systems is a characteristic feature of
quantum mechanics which plays a key role in the foundations of the
theory. Moreover, it is a fundamental resource in many of the
applications of quantum information theory \cite{Nie}. Therefore,
the characterization of entangled states (the so-called separability
problem) is of great interest, since a complete understanding of
their structure would lead to a better understanding of quantum
theory and it would also clarify which states are useful for quantum
information tasks (e.g., in an experimental context). However, and
despite many efforts in the last decade, the separability problem
remains unsolved. In fact, it has been proved to be NP-hard
\cite{NP}, although several authors have devised non-trivial
algorithms for it (see \cite{Ioa} for a survey). Nevertheless, there
exist a great variety of analytical approaches to the problem which,
besides their entanglement detection capability, give as a
by-product a better insight in the structure of entangled states
from the theoretical point of view. Historically, Bell inequalities
were the first tool for the recognition of entanglement; however, it
is well-known for some time that the violation of a Bell inequality
is only a sufficient condition for entanglement and not a necessary
one, and that there are in fact many entangled states that satisfy
them (see, e.g., \cite{Wer}). The most remarkable sufficient
condition for the detection of entanglement is given by the
Peres-Horodecki or positive partial transpose (PPT) criterion
\cite{Per}. This condition is so strong that it characterizes
entanglement for $2\times2$ and $2\times3$ systems \cite{Hor1}, but
not otherwise. Much subsequent work has been devoted to finding
different sufficient conditions for the presence of entanglement
that could complement the strong Peres-Horodecki criterion by
detecting PPT entanglement (i.e., entangled states which are not
detected by this criterion). The most remarkable one is the
computable cross norm \cite{Rud} or realignment \cite{Che} (CCNR)
criterion, which exhibits a powerful PPT entanglement detection
capability. There are also other important criteria with this
property, which, however, lack the operational character (i.e., ease
of use) of the aforementioned ones because they rely on expectation
values of observables which have to be chosen appropriately for the
state in question. This is the case of conditions based on
entanglement witnesses (see e.g. \cite{Hor1,EW}) or uncertainty
relations \cite{ur}.

In a recent paper \cite{deV1}, a new easily computable operational
sufficient condition for entanglement in bipartite quantum systems
of arbitrary dimensions $M\times N$ has been obtained by studying
separability from the point of view of the Bloch representation of
density matrices. We will refer to it as correlation matrix (CM)
criterion. It was proved that it is able to recognize PPT
entanglement when $M=N$, and that, although strictly weaker than the
CCNR criterion when $M=N$ (i.e., the CCNR criterion detects all
entangled states recognized by the CM criterion in this situation
and not conversely), it can be stronger when $M\neq N$. Therefore,
it remained as an open question whether the CM criterion could be
stronger than both the PPT and CCNR criteria for certain states. We
will show in this paper that this is indeed the case by providing
examples of PPT entangled states not detected by the CCNR criterion
whose entanglement is revealed by the CM criterion. This result
further confirms the interest of the CM criterion, showing that it
actually complements the PPT and CCNR criteria improving our ability
to detect entanglement. We will also consider a very recently
derived criterion \cite{CMC}, which is strictly stronger than the
CCNR.

In the last part of this paper we will consider the possibility of
quantifying entanglement through the CM. Besides the
characterization of entanglement, the derivation of good
entanglement measures is a fundamental problem in quantum
information theory since they provide a way to quantify how useful
an entangled state is for a certain quantum information process.
Several measures of this type exist (see, e.g., the recent survey
\cite{Ple}), but, in general, their explicit computation is a very
hard task, even numerically. Therefore, it is interesting to find
good bounds for them (see \cite{Ple,Buc} and references therein). It
is widely believed that a stronger violation of a separability
condition indicates a bigger amount of entanglement, and, in fact it
has been shown that they can be used to place lower bounds on
different entanglement measures \cite{caf1,caf2,deV2} (see also
\cite{Dat} for bounds obtained by considering two separability
conditions simultaneously). In particular, the measure known as
concurrence can be bounded from below using the PPT and CCNR
criteria \cite{caf1}, and the CM criterion \cite{deV2}. These
results can be used to obtain lower bounds for its close cousin the
tangle. However, we will prove here that the CM can be used to
obtain different lower bounds for this measure, which are
particularly sharp for states of high purity. This follows from the
fact that the CM provides the actual value of both tangle and
concurrence for pure states as we will also show. Finally, as an
application, we will use this new bound to prove a recently
conjectured result relating concurrence and the so-called
Minkowskian-norm-based (MNB) entanglement measure for two-qubit
systems ($M=N=2$) \cite{mink1,mink2}.

\section{Separability, Bloch representation and CM criterion}

Consider a bipartite quantum system composed of subsystems $A$ and
$B$, of $M$ and $N$ levels respectively. Then, its quantum state is
characterized by the density operator $\rho\in
\mathcal{B}(H_A\otimes H_B)$, where $H_A\simeq\mathbb{C}^M$ and
$H_B\simeq\mathbb{C}^N$ denote the Hilbert spaces of the subsystems
and $\mathcal{B}(H)$ stands for the real vector space of Hermitian
operators acting on $H$ with the standard Hilbert-Schmidt inner
product $\langle\rho,\tau\rangle_{HS}=\textrm{Tr}(\rho^\dag\tau)$.
The state is said to be separable (entangled) if it can (cannot) be
written as a convex combination of product states \cite{Wer}, i. e.
\begin{equation}\label{separable}
\rho=\sum_i p_i \, \rho^A_i\otimes\rho^B_i,
\end{equation}
where $0 \leq p_i \leq 1$, $\sum_i p_i = 1$, and $\rho_i^A$
($\rho_i^B$) denotes a pure state density matrix acting on $H_A$
($H_B$). Given that it is in general very hard to verify if a
decomposition according to the definition of separability
(\ref{separable}) exists for a given state, the separability problem
consists in finding computable mathematical conditions which provide
a practical way to check whether a given state is entangled or not.
The PPT and CCNR criteria can be formulated in several ways. They
can be easily applied by considering certain rearrangements of the
matrix elements of $\rho$ \cite{Hor2}. For instance, if we expand
with respect to the canonical product basis
\begin{equation}
\rho=\sum_{ijkl} \rho_{ij,kl}|ij\rangle\langle kl|,
\end{equation}
the PPT criterion states that for separable states the partial
transpose $T_A(\rho)$, i.e.
\begin{equation}\label{ppt}
T_A(\rho)_{ij,kl}=\rho_{kj,il},
\end{equation}
still represents a state and it is, therefore, positive
semidefinite, which means that $||T_A(\rho)||_{tr}=1$, where here
and throughout the paper $||\cdot||_{tr}$ stands for the trace or Ky
Fan norm (i.e. the sum of the singular values). Similarly, the CCNR
criterion affirms that the realignment operation $R(\rho)$,
\begin{equation}\label{realignment}
R(\rho)_{ij,kl}=\rho_{ik,jl},
\end{equation}
is such that $||R(\rho)||_{tr}\leq1$ for all separable states.
Hence, $||T_A(\rho)||_{tr},||R(\rho)||_{tr}>1$ is a signature of
entanglement.

To introduce the CM criterion we first say a few words about the
Bloch representation of density matrices \cite{Blo,Hio}, which is an
expansion of $\rho$ in terms of the orthogonal basis of
$\mathcal{B}(H)$ given by the identity $I$ and the traceless
Hermitian generators of the group $SU(\dim H)$ $\{\lambda_i\}$,
\begin{equation}\label{bloch}
\rho=\frac{1}{\dim H}\left(I+\sum_{i=1}^{(\dim
H)^2-1}r_i\lambda_i\right).
\end{equation}
The generators fulfill the following orthogonality relations (the
normalization is chosen by standard usage),
\begin{equation}\label{normalizacion}
\langle\lambda_i,I\rangle_{HS}=\textrm{Tr}(\lambda_i)=0,\quad
\langle\lambda_i,\lambda_j\rangle_{HS}=\textrm{Tr}(\lambda_i\lambda_j)=2\delta_{ij},
\end{equation}
and they can be easily constructed from any orthonormal basis in $H$
\cite{Hio}. The coefficients $\{r_i\}$ which completely characterize
$\rho$ form the coherence or Bloch vector
$\textbf{r}\in\mathbb{R}^{(\dim H)^2-1}$. The orthogonality of the
basis implies that this vector can be easily determined
(theoretically and experimentally) from the expectation values of
the observables $\{\lambda_i\}$
\begin{equation}\label{ri}
r_i=\frac{\dim H}{2}\langle\rho,\lambda_i\rangle_{HS}=\frac{\dim
H}{2}\textrm{Tr}(\rho\lambda_i)=\frac{\dim
H}{2}\langle\lambda_i\rangle_\rho.
\end{equation}
In the case of $M\times N$ bipartite quantum systems ($M\leq N$ is
assumed without loss of generality throughout the paper), the Bloch
representation (also known as Fano form \cite{Fan}) can be written
as
%\begin{align}\label{bipartitebloch}
%\rho =& \frac{1}{MN}\left(I_M\otimes I_N+\sum_ir_i\lambda^A_i\otimes
%I_N+\sum_js_jI_M\otimes\lambda^B_j\right.\nonumber\\
%& \left.+\sum_{i,j}t_{ij}\lambda^A_i\otimes\lambda^B_j\right),
%\end{align}
\begin{equation}\label{bipartitebloch}
\fl \rho = \frac{1}{MN}\left(I_M\otimes
I_N+\sum_ir_i\lambda^A_i\otimes
I_N+\sum_js_jI_M\otimes\lambda^B_j+\sum_{i,j}t_{ij}\lambda^A_i\otimes\lambda^B_j\right),
\end{equation}
where $\{\lambda_i^A\}_{i=1}^{M^2-1}$ and
$\{\lambda_i^B\}_{i=1}^{N^2-1}$ denote the generators of $SU(M)$ and
$SU(N)$. This representation has to kind of parameters:
$\{r_i\}=M/2\{\langle\lambda_i^A\otimes I_N\rangle_\rho\}$ and
$\{s_i\}=N/2\{\langle I_M\otimes\lambda_j^B\rangle_\rho\}$, which
are local since they are the Bloch parameters of the reductions
($\rho_A=\textrm{Tr}_B\rho=1/M(I+\sum_ir_i\lambda_i^A)$,
$\rho_B=\textrm{Tr}_A\rho=1/N(I+\sum_is_i\lambda^B_i)$); and
$\{t_{ij}\}=MN/4\{\langle\lambda_i^A\otimes\lambda_j^B\rangle_\rho\}$,
which are responsible for the possible correlations between the
subsystems and form the CM $T\in\mathbb{R}^{(M^2-1)\times (N^2-1)}$.
The CM criterion states that
\begin{equation}\label{CM}
||T||_{tr}\leq\sqrt{\frac{MN(M-1)(N-1)}{4}}
\end{equation}
must hold for all separable states \cite{deV1}. This implies that
there is an upper bound to the ``amount'' of correlations contained
in this kind of states, a higher degree of correlations only being
achievable through entanglement.

\section{Entangled states detected by the CM criterion but not by
the PPT and CCNR criteria}

In \cite{deV1} it is proved that the CCNR criterion is stronger than
the CM criterion when $M=N$. However, in the case of states with
maximally mixed subsystems (i.e., \textbf{r}=\textbf{s}=0) it is
shown that the CM criterion is strictly stronger than the CCNR
criterion when $M\neq N$, being equivalent when $M=N$. Several
examples of $M=N$ PPT entangled states detected by condition
(\ref{CM}) are also provided. However, since the CCNR condition is
stronger in this case, the entanglement of all these states is
already revealed by this criterion. Therefore, to find examples of
entangled states just detected by the CM criterion we have to
restrict ourselves to the $M\neq N$ case. Unfortunately, although
many examples of PPT entangled states are known when $M=N$ (see,
e.g., \cite{PHor,upb1,upb2}), the situation in the asymmetric case
is not as rich. In \cite{PHor} a set of PPT entangled states in
$2\times4$ dimensions is provided (see the paper for their explicit
form). However, it can be readily checked that both the CCNR and CM
criteria are unable to identify these states as entangled as well.
In \cite{upb1}, an entangled PPT state with subsystems of different
dimensions ($M,N\geq3$) is constructed from the unextendible product
basis GenTiles2 (its explicit form is given below). However, while
the CCNR criterion detects entanglement for this state when $M=3$,
$4\leq N\leq8$ and when $M=4$, $N=5$, the CM criterion only
recognizes entanglement in the $3\times4$ case. So, unfortunately
for our purpose, the CCNR criterion seems stronger for these states.
The entanglement properties of rotationally invariant states has
been thoroughly studied in recent literature. In particular, the set
of PPT rotationally invariant states has been determined in $3\times
N$ \cite{Bre2} and $4\times N$ \cite{Aug} systems, it has been shown
that the Peres-Horodecki criterion does not characterize
entanglement in this scenario and many PPT entangled states have
been identified. Although this situation may seem promising to our
purpose since rotationally invariant states have maximally
disordered subsystems and, therefore, the CM criterion is guaranteed
to improve on the CCNR criterion, our numerical explorations
indicate that the PPT criterion is stronger than the CM criterion
for these states.

The above states are, to our knowledge, the only examples available
in the literature of PPT entanglement with subsystems of different
dimensions. Therefore, to achieve our goal we have to either
construct new examples or to consider PPT-preserving operations on
the previous states which transform them to different states in
which the CM criterion is stronger than the CCNR, such as the
$\textbf{r}=\textbf{s}=0$ case. It turns out that this
transformations are very well studied. In \cite{normalform} (see
also \cite{Lei}) it is shown that every full-rank state can be
transformed under the action of local filtering operations (also
known as stochastic local operations assisted by classical
communication SLOCC) into a state with maximally mixed subsystems
which is called (filter) normal form (FNF). This form is unique up
to local unitary transformations. Mathematically, the filtering
operations are represented by invertible matrices $F_A, F_B$
(without loss of generality they can be chosen of determinant equal
to one) which transform $\rho$ into $\tilde{\rho}$ according to
\begin{equation}\label{FNF}
\tilde{\rho}=\frac{(F_A\otimes F_B)\rho(F_A\otimes
F_B)^{\dag}}{\textrm{Tr}(F_A\otimes F_B)\rho(F_A\otimes
F_B)^{\dag}}.
\end{equation}
There exist several constructive algorithms which provide the
matrices $F_A, F_B$ needed in order to take a given state into its
normal form \cite{normalform,Lei} (see also \cite{CMC}). In
particular, there is a matlab code available in \cite{Ver}. Although
the existence of the FNF is guaranteed only for full-rank density
matrices, rank-deficient states can be brought to a state whose
reductions are arbitrarily close to the maximally mixed state using
the aforementioned algorithms. The transformation given by
(\ref{FNF}) preserves the separability or entanglement of a given
state as well as the PPT property, so the CM criterion will be
stronger than the PPT and CCNR criteria for states $\tilde{\rho}$ in
the FNF obtained from a PPT entangled state. The FNF was first
considered in the context of the separability problem in \cite{CMC}:
since these SLOCC transformations that wash out all the local
information maximize the entanglement content of a state in a
certain sense \cite{normalform}, the entanglement detection
capability of separability conditions is greatly improved in the
FNF. However, here we will just consider it as a way to construct a
state with some desired properties.

We start by considering the entangled PPT state $\rho_{GT2}$
constructed from the unextendible product basis GenTiles2
\cite{upb2,upb3} in $M\times N$ dimensions such that $M\geq3$, $N>3$
and $M\leq N$,
\begin{equation}\label{gt2}
\fl \rho_{GT2} = \frac{1}{2M-1}\left(I_{MN}-|F\rangle\langle
F|-\sum_{j=0}^{M-1}|S_j\rangle\langle
S_j|-\sum_{j=0}^{M-1}\sum_{k=1}^{N-3}|L_{jk}\rangle\langle
L_{jk}|\right),
\end{equation}
%\begin{align}\label{gt2}
%\rho =& \frac{1}{2M-1}\left(I_{MN}-|F\rangle\langle
%F|-\sum_{j=0}^{M-1}|S_j\rangle\langle
%S_j|\right.\nonumber\\
%& \left.-\sum_{j=0}^{M-1}\sum_{k=1}^{N-3}|L_{jk}\rangle\langle
%L_{jk}|\right),
%\end{align}
where
\begin{eqnarray}
\eqalign{
|F\rangle &= \frac{1}{\sqrt{MN}}\sum_{i=0}^{M-1}\sum_{j=0}^{N-1}|ij\rangle,\\
|S_j\rangle &= \frac{1}{\sqrt{2}}(|j\rangle-|j+1 \textrm{ mod }
M\rangle)|j\rangle,\\
|L_{jk}\rangle &=
\frac{1}{\sqrt{N-2}}|j\rangle\left(\sum_{l=0}^{M-3}\exp\left(\rmi\frac{2\pi
lk}{N-2}\right)|l+j+1 \textrm{ mod }
M\rangle\right.\\
&+\left.\sum_{l=M-2}^{N-3}\exp\left(\rmi\frac{2\pi
lk}{N-2}\right)|l+2\rangle\right).}
\end{eqnarray}
For simplicity we restrict ourselves to the lowest possible
dimensions $M=3$ and $N=4$. In this case
$||T_{\rho_{GT2}}||_{tr}=4.3428$ and $||R(\rho_{GT2})||_{tr}=1.0315$
and, therefore, as said before, both the CM and CCNR criteria detect
$\rho_{GT2}$ as entangled. It can be seen by considering any of the
algorithms mentioned above that the filters
\begin{eqnarray}
\fl F_A =\left(
      \begin{array}{rrr}
        -0.2586 - 0.4251\rmi & -0.2586 - 0.4251\rmi & -0.2586 - 0.4251\rmi \\
        0.3421 - 0.3842\rmi & 0.4402 + 0.2817\rmi & -0.7824 + 0.1025\rmi \\
        0.2784 - 0.6568\rmi & -0.5774 + 0.4086\rmi & 0.2990 + 0.2482\rmi \\
      \end{array}
    \right),\nonumber\\
\fl F_B =\left(
      \begin{array}{rrrr}
        -0.3118 - 0.3092\rmi & -0.3118 - 0.3092\rmi & -0.3118 - 0.3092\rmi & -0.3118 - 0.3092\rmi \\
        0.5499 - 0.2805\rmi & 0.6414 - 0.0813\rmi & -0.3307 + 0.0334\rmi & -0.4303 + 0.1642\rmi \\
        -0.3932 - 0.1066\rmi & 0.3198 - 0.3909\rmi & -0.0427 - 0.7619\rmi & 0.0580 + 0.6297\rmi \\
        0.5358 + 0.3605\rmi & 0.1113 - 0.5279\rmi & 0.5169 - 0.0640\rmi & -0.5820 + 0.1157\rmi \\
      \end{array}
    \right),\nonumber
\end{eqnarray}
transform $\rho_{GT2}$ onto its normal form $\tilde{\rho}_{GT2}$.
Now, we readily find that $||T_{\tilde{\rho}_{GT2}}||_{tr}=4.5751$
and $||R(\tilde{\rho}_{GT2})||_{tr}=1.0512$, and again both the CM
and CCNR criteria reveal the entanglement of this PPT state.
However, the CM criterion is now stronger and, therefore, more
robust against noise. So if we consider the previous state mixed
with white noise, i.e.
\begin{equation}
\rho(p)=p\tilde{\rho}_{GT2}+(1-p)\frac{I_{12}}{12},
\end{equation}
we find that the CM criterion detects entanglement in $\rho(p)$
whenever $p\geq0.9274$ while the CCNR criterion recognizes
entanglement when $p\geq0.9330$. Notice that by construction the PPT
criterion is unable to find entanglement in $\rho(p)$. Thus, this
example shows that the CM criterion can detect states which are
neither detected by the CCNR criterion nor by the PPT criterion.
Moreover, the recently powerful criterion based on covariance
matrices derived in \cite{CMC} (which is strictly stronger than the
CCNR) detects entanglement in this state when $p\geq0.9290$ and it
is, therefore, also weaker than the CM criterion for these states,
as was expected, since in the above mentioned paper it is shown that
the new criterion is stronger than the CM criterion when $M\ll N$
but otherwise weaker in the case of states with maximally disordered
subsystems. Several other examples of entangled states detected by
the CM criterion but not by the PPT and CCNR criteria can be found
considering the FNF of $\rho_{GT2}$ for other values of $M$ and $N$.
However, it is worth pointing out, that the FNF of the $2\times4$
states of \cite{PHor} is still undetected by the CM and CCNR
criteria and that, in fact,
$||T_{\tilde{\rho}}||_{tr}<||T_{\rho}||_{tr}$ and
$||R(\tilde{\rho})||_{tr}<||R(\rho)||_{tr}$ for many of these
states.

\section{Tangle and the CM}

The entanglement of formation \cite{EOF} is the only measure of
entanglement for which an analytical expression is available for
arbitrary systems of particular dimensions. It was found in
\cite{Woo} for the case of two-qubit systems. In this case the
entanglement of formation is a monotonically increasing function of
a quantity called concurrence C, so C is taken as a measure of
entanglement in its own right. Furthermore, it has been successfully
generalized to arbitrary dimensional bipartite quantum states
\cite{Run1,Run2} (see also \cite{Alb}). For a pure state $\psi$, it
is given by \cite{Run1}
\begin{equation}\label{pureconcurrence}
C(\psi)=\sqrt{2(1-\textrm{Tr}\rho_A^2)}.
\end{equation}
Notice that $0\leq C(\psi)\leq \sqrt{2(M-1)/M}$, the lower bound
being attained by product states and the upper bound by maximally
entangled states. The definition is extended to general mixed states
$\rho$ by the convex roof (the minimum average value of the
pure-state measure over all possible ensemble realizations of
$\rho$) \cite{Run2},
\begin{equation}\label{concurrence}
C(\rho)=\min_{\{p_i,|\psi_i\rangle\}}\left\{\sum_ip_iC(\psi_i) :
\rho=\sum_ip_i|\psi_i\rangle\langle\psi_i|\right\}.
\end{equation}
Consequently, $C(\rho)=0$ if, and only if, $\rho$ is a separable
state. It can be more convenient to remove the square root in
(\ref{pureconcurrence}) and consider the measure
$\tau(\psi)=C^2(\psi)$, which is then extended to mixed states by
the convex roof
\begin{equation}\label{tangle}
\tau(\rho)=\min_{\{p_i,|\psi_i\rangle\}}\left\{\sum_ip_iC^2(\psi_i)
: \rho=\sum_ip_i|\psi_i\rangle\langle\psi_i|\right\}.
\end{equation}
The measure $\tau$ is known as tangle. Notice that, although equal
to the squared concurrence for pure states, for general states it
holds that $\tau(\rho)\geq C^2(\rho)$; nevertheless, it can be shown
that $\tau(\rho)=C^2(\rho)$ in the case of two-qubit states (see
\cite{Osb}). Some authors have found the tangle a more natural
measure than the concurrence because a closed formula can be derived
for it for rank-2 density operators \cite{Osb} and because, contrary
to the concurrence, its behaviour is analogous to the entanglement
of formation for isotropic states \cite{Run2}.

Due to the convex roof construction these measures are very hard to
compute in the case of mixed states, so, as mentioned in Sec.\ I,
good bounds for their estimation are desirable. In particular, lower
bounds are preferable because upper bounds can be obtained
considering any ensemble decomposition of the state. It seems
natural to think that $||T_A(\rho)||_{tr}$ and $||R(\rho)||_{tr}$
provide an estimate of the entanglement content of $\rho$ since the
greater than 1 they are, the further the state is to separability in
a certain sense. In fact, it has been proved that \cite{caf1}
\begin{equation}\label{caf}
C(\rho)\geq\sqrt{\frac{2}{M(M-1)}}\left[\max(||T_A(\rho)||_{tr},||R(\rho)||_{tr})-1\right],
\end{equation}
which provides a powerful lower bound to estimate the concurrence
from these separability conditions. It has been shown in \cite{deV2}
that the CM can be used analogously since
\begin{equation}\label{CMbound}
C(\rho)\geq\sqrt{\frac{8}{M^3N^2(M-1)}}\left(||T||_{tr}-\sqrt{\frac{MN(M-1)(N-1)}{4}}\right).
\end{equation}
The bound given by (\ref{caf}) is generally tighter than that of
(\ref{CMbound}) (see \cite{deV2}). However, using the results of
Sec. II we can provide examples of the contrary (which lacked in
\cite{deV2}). For instance, while (\ref{caf}) tells us that
$C(\tilde{\rho}_{GT2})\geq0.0296$, we have that
$C(\tilde{\rho}_{GT2})\geq0.0320$ using (\ref{CMbound}). The above
formulas can also be used for the tangle recalling that
$\tau(\rho)\geq C^2(\rho)$. However, here we will derive a different
lower bound which is exclusively designed for the tangle which is
better than Eqs.\ (\ref{caf})-(\ref{CMbound}) in certain situations.

The matrix $T$ contains the information about the correlations
between the subsystems and the CM criterion bounds the amount of
correlations in a separable state using the trace norm to quantify
them. However, this criterion can be stated using any matrix norm
since the proof only relies on the triangle inequality \cite{deV1}.
As we shall discuss in more detail below, the choice of the trace
norm is convenient because it provides the strongest separability
condition; nevertheless, different choices can be more adequate if
we are interested in the quantification of entanglement. This is the
underlying idea for the new bound on the tangle to be derived in
this Section. We first show that $\tau$ and $C$ are closely related
to the CM for pure states and, moreover, that they can be evaluated
by considering a particular norm of the CM.

\subsection{Pure states}

The concurrence and tangle of a pure state given by
(\ref{pureconcurrence}) can be easily written in terms of the
parameters of the Bloch representation. Recall that the reduced
density matrix of of an arbitrary state $\rho$ with Bloch
representation (\ref{bipartitebloch}) is
$\rho_A=1/M(I+\sum_ir_i\lambda_i^A)$. Thus, using
(\ref{normalizacion}), it can be seen that
\begin{equation}\label{trazareduccion}
\textrm{Tr}(\rho^2_A)=\frac{M+2||\textbf{r}||^2_2}{M^2},
\end{equation}
where $||\cdot||_2$ is the Euclidean norm. Hence,
\begin{equation}\label{tanglerpure}
\tau(\psi)=C^2(\psi)=\frac{2(M^2-M-2||\textbf{r}||_2^2)}{M^2}.
\end{equation}

In the case of pure states $\textbf{r}$, $\textbf{s}$ and $T$ are
related in a determined way. For these states the reductions
$\rho_A$ and $\rho_B$ have the same eigenvalues and, therefore,
$\textrm{Tr}(\rho^2_A)=\textrm{Tr}(\rho^2_B)$. So, recalling
(\ref{trazareduccion}), this implies that
\begin{equation}\label{rs}
\frac{M+2||\textbf{r}||^2_2}{M^2}=\frac{N+2||\textbf{s}||^2_2}{N^2}.
\end{equation}
Furthermore, pure states satisfy $\textrm{Tr}(\rho^2)=1$. Using
again (\ref{normalizacion}) and some straightforward algebra, we
readily see that this means that
\begin{equation}\label{rst}
N||\textbf{r}||^2_2+M||\textbf{s}||_2^2+2||T||_{HS}^2=\frac{MN(MN-1)}{2},
\end{equation}
where $||\cdot||_{HS}$ is the Frobenius or Hilbert-Schmidt norm,
that is, the norm induced by the Hilbert-Schmidt inner product, i.e.
\begin{equation}\label{HS}
||T||_{HS}=\sqrt{\textrm{Tr}(T^\dag T)}=\sqrt{\sum_{ij}|t_{ij}|^2}.
\end{equation}
Eqs. (\ref{rs}) and (\ref{rst}) imply that the value of one of the
parameters $\{||\textbf{r}||_2,||\textbf{s}||_2,||T||_{HS}\}$
uniquely determines the others in the case of pure states. Thus,
inserting (\ref{rs}) in (\ref{rst}) we arrive at
\begin{equation}
||\textbf{r}||^2_2=\frac{M}{M+N}\left(\frac{N(M^2-1)}{2}-\frac{2}{N}||T||_{HS}^2\right).
\end{equation}
Now, this last equation together with (\ref{tanglerpure}) lets us
write the concurrence and tangle of an arbitrary pure bipartite
state in terms of the CM,
\begin{equation}\label{tangleCMpure}
\tau(\psi)=C^2(\psi)=\frac{8}{MN(M+N)}\left(||T||^2_{HS}-\frac{MN(M-1)(N-1)}{4}\right).
\end{equation}
In this way we see that for pure states, the CM not only
characterizes entanglement but also enables to rigorously quantify
it, since concurrence and tangle are functions of $||T||_{HS}$.

\subsection{Mixed states}

In the case of mixed states $||T||_{HS}$ cannot be used to express
the tangle or concurrence in closed form as in (\ref{tangleCMpure}).
However, if we restrict ourselves to the tangle, it is possible to
derive a lower bound for this measure similar to (\ref{CMbound})
using $||T||_{HS}$. Let $\sum_np_n|\psi_n\rangle\langle\psi_n|$ be
the decomposition of $\rho$ for which the minimum in (\ref{tangle})
is attained. Then, we have that
\begin{eqnarray}
\tau(\rho)&=\sum_np_n\tau(\psi_n)\nonumber\\
&=\frac{8}{MN(M+N)}\left(\sum_np_n||T_{\psi_n}||^2_{HS}-\frac{MN(M-1)(N-1)}{4}\right)\nonumber\\
&\geq\frac{8}{MN(M+N)}\left(\left(\sum_np_n||T_{\psi_n}||_{HS}\right)^2-\frac{MN(M-1)(N-1)}{4}\right)\nonumber\\
&\geq\frac{8}{MN(M+N)}\left(\left|\left|\sum_np_nT_{\psi_n}\right|\right|_{HS}^2-\frac{MN(M-1)(N-1)}{4}\right)\nonumber\\
&=\frac{8}{MN(M+N)}\left(||T_\rho||^2_{HS}-\frac{MN(M-1)(N-1)}{4}\right),\label{newbound}
\end{eqnarray}
where in the first inequality we have used the convexity of the
function $f(x)=x^2$ and in the second inequality we have used the
convexity of $||\cdot||_{HS}$ (i.e., the triangle inequality).
Equation (\ref{newbound}) implies the following separability
criterion: $||T||_{HS}\leq\sqrt{MN(M-1)(N-1)}/2$ holds for all
separable states. However, this condition is weaker than the CM
criterion and can be trivially deduced from it since
$||T||_{tr}\geq||T||_{HS}$. Therefore, (\ref{newbound}) will not
place new non-trivial bounds where (\ref{caf}) and (\ref{CMbound})
failed, moreover, the latter equations place non-trivial bounds
where the former fails. However, this new bound yields the exact
value of the tangle of pure states while the others do not. Thus,
although weaker for the detection of entanglement, it will be
tighter for the estimation of the tangle for states which are close
to pure states. This kind of states are common in experiments where
the pure entangled state $\psi$ one aims to prepare is subjected to
different types of noise. As a result a slightly mixed state is
finally obtained:
$\rho=p|\psi\rangle\langle\psi|+(1-p)\rho_{\textrm{noise}}$ with $p$
close to 1. To test the new bound (\ref{newbound}), we have
considered mixtures in $3\times3$ dimensions of arbitrary pure
entangled states and white noise (i.e.
$\rho_{\textrm{noise}}=I_9/9$). We have found that, in general, the
bound of (\ref{newbound}) can be better than Eqs.\
(\ref{caf})-(\ref{CMbound}) when $p\gtrsim0.94$.

It is also worth pointing out that using a similar reasoning, it is
possible to derive an upper bound for the tangle in terms of
$||\textbf{r}||_2$ and $||\textbf{s}||_2$ if we start from
(\ref{tanglerpure}) (or its equivalent in terms of $N$ and
$||\textbf{s}||_2$),
\begin{equation}\label{upperbound}
\tau(\rho)\leq2\min\left\{\frac{M^2-M-2||\textbf{r}||_2^2}{M^2},\frac{N^2-N-2||\textbf{s}||_2^2}{N^2}\right\}.
\end{equation}

\subsection{Concurrence and MNB entanglement measure
for two-qubit systems}

The MNB measure is an entanglement measure for two-qubit states
which is defined as \cite{mink1}
\begin{equation}\label{mink}
E(\rho)=\max\{\tr\rho^2-1+\tr\rho(\sigma_y\otimes\sigma_y)\rho^*(\sigma_y\otimes\sigma_y),0\},
\end{equation}
where $\{\sigma_x,\sigma_y,\sigma_z\}$ denote the standard Pauli
matrices (i.e., the generators of $SU(2)$) and $\rho^*$ denotes
complex conjugation of the density matrix, which is taken, as usual,
in the basis of eigenstates of $\sigma_z$. Although $E(\rho)=0$ for
many entangled states, this measure can be analytically computed and
it has been shown to be a rigorous entanglement measure (i.e.\
non-increasing on average under LOCC) in \cite{mink2}. In this same
paper it is shown that $E(\rho)$ lower bounds $C(\rho)$ for
particular sets of states. Based on numerical evidence the authors
suggest that this could hold for all states. As an application of
our results, we will use the bound (\ref{newbound}) to prove this
conjecture.

Using equations (5) in \cite{mink2} and (\ref{HS}) here it is
clearly seen that the MNB measure can be written as
\begin{equation}
E(\rho)=\max\{\frac{1}{2}\left(||T||_{HS}^2-1\right),0\},
\end{equation}
which is precisely the bound for the tangle obtained in
(\ref{newbound}) for two-qubit states. Hence, we have that
$\tau(\rho)\geq E(\rho)$. On the other hand, since for two qubits we
have that $0\leq C(\rho)\leq1$ it holds that $C(\rho)\geq
C^2(\rho)$. Now, recalling that in this case $\tau(\rho)=C^2(\rho)$
we arrive at the desired result: $C(\rho)\geq E(\rho)$. Thus, we see
that the MNB entanglement measure is directly related to the CM and
that it lower bounds both the concurrence and tangle.

\section{Conclusions}

The CM criterion provides a general operational sufficient condition
for entanglement which, besides its theoretical interest in the
theory of entanglement, offers a relatively simple scheme for its
detection in experiments. While other important criteria of this
type, such as the PPT and CCNR criteria, demand full knowledge of
the density operator ($M^2N^2-1$ parameters need to be specified),
the CM depends on $(M^2-1)(N^2-1)$ parameters to be determined by
measurements of local operators
($t_{ij}=MN\langle\lambda_i^A\otimes\lambda_j^B\rangle_\rho/4$) and,
hence, the CM criterion requires in principle less experimental
effort. Furthermore, it relies on a measure, $||T||_{tr}$, which is
left invariant under local unitary transformations of the density
operator \cite{deV1} and, therefore, the measurement setups of $A$
and $B$ need not be aligned which, contrary to entanglement tests
based on uncertainty relations, also eases its experimental
implementation (see \cite{Kot} for a more detailed discussion).

In this paper we have further confirmed the interest of the CM
criterion by providing examples of entangled states which are not
detected by the PPT and CCNR criteria whose entanglement is
identified by this criterion. Thus, it is clear that the CM
criterion together with the previous criteria improves our ability
to characterize entanglement, although there remain entangled states
unrevealed by the three criteria and the characterization is,
therefore, not complete.

Like the PPT and CCNR criteria, the CM criterion can be used to
estimate the entanglement content of a state. Here, we have also
studied the possibility of quantifying the correlations inherent in
the CM by considering a different norm than the one used in the
separability criterion. We have found that, although the trace norm
is more suitable for the detection of entanglement, the
Hilbert-Schmidt norm of the CM is also appropriate for the
quantification of entanglement. In fact, it constitutes a rigorous
entanglement measure for pure states, given that the concurrence and
tangle are monotonously increasing functions of it. Despite that
this relation does not hold for mixed states we have shown that it
provides a lower bound for the tangle, particularly sharp for states
of high purity, that can actually improve the estimations given by
the PPT, CCNR and CM criteria. Therefore, the determination of the
CM also improves our ability to estimate entanglement measures.

\begin{ack}

The author thanks O. G\"uhne for discussions on filter normal forms.
Financial support by Universidad Carlos III de Madrid and Comunidad
Aut\'onoma de Madrid (Project No.\ CCG06-UC3M/ESP-0690) and by
Direcci\'on General de Investigaci\'on (Ministerio de Educaci\'on y
Ciencia) under Grant No.\ MTM2006-13000-C03-02 is gratefully
acknowledged.

\end{ack}


\section*{References}



\begin{thebibliography}{widest-label}

\bibitem{Nie} Nielsen M A and Chuang I L 2000 \textit{Quantum
Computation and Quantum Information} (Cambridge University Press)

\bibitem{NP} Gurvits L 2003 \textit{Proc. of the
35 Annual ACM Symp. Theory of Computing (San Diego)} (New York: ACM
Press) p 10

\bibitem{Ioa} Ioannou L M 2007 \textit{Quantum Inf. Comput.} \textbf{7} 335

\bibitem{Wer} Werner R F 1989 \textit{Phys. Rev. A} \textbf{40} 4277

\bibitem{Per} Peres A 1996 \textit{Phys. Rev. Lett.} \textbf{77} 1413

\bibitem{Hor1} Horodecki M, Horodecki P and Horodecki R 1996 \textit{Phys. Lett. A}
\textbf{223} 1

\bibitem{Rud} Rudolph O 2002 Further results on the cross norm criterion for separability
\textit{Preprint} arXiv:quant-ph/0202121v1

\bibitem{Che} Chen K and Wu L A 2003 \textit{Quantum Inf. Comput.} \textbf{3} 193

\bibitem{EW} Lewenstein M, Kraus B, Cirac J I and Horodecki P 2000 \textit{Phys.
Rev. A} \textbf{62} 052310

\bibitem{ur} Hofmann H F and Takeuchi S 2003 \textit{Phys. Rev. A} \textbf{68} 032103

\nonum G\"{u}hne O 2004 \textit{Phys. Rev. Lett.} \textbf{92} 117903

\nonum G\"uhne O, Mechler M, T\'{o}th G and Adam P 2006
\textit{Phys. Rev. A} \textbf{74} 010301(R)

\bibitem{deV1} de Vicente J I 2007 \textit{Quantum Inf. Comput.} \textbf{7} 624

\bibitem{CMC} G\"{u}hne O, Hyllus P, Gittsovich O and Eisert J 2007 \textit{Phys. Rev. Lett.} \textbf{99}
130504

\bibitem{Ple} Plenio M B and Virmani S 2007 \textit{Quantum Inf. Comput.} \textbf{7} 1

\bibitem{Buc} Mintert F, Carvalho A R R, Ku\'s M and Buchleitner A
2005 \textit{Phys. Rep.} \textbf{415} 207

\bibitem{caf1} Chen K, Albeverio S and Fei S M 2005 \textit{Phys. Rev. Lett.} \textbf{95}
040504

\bibitem{caf2} Chen K, Albeverio S and Fei SM 2005 \textit{Phys. Rev. Lett.} \textbf{95}
210501

\nonum Breuer H P 2006 \textit{J. Phys. A: Math. Gen.} \textbf{39}
11847

\nonum Audenaert K M R and Plenio M B 2006 \textit{New J. Phys.}
\textbf{8} 266

\nonum Eisert J, Brand\~{a}o F G S L and Audenaert K M R 2007
\textit{New J. Phys.} \textbf{9} 46

\nonum G\"uhne O, Reimpell M and Werner R F 2007 \textit{Phys. Rev.
Lett.} \textbf{98} 110502

\nonum Mintert F 2007 \textit{Phys. Rev. A} \textbf{75} 052302

\nonum Zhang C J, Zhang Y S, Zhang S and Guo G C 2007 \textit{Phys.
Rev. A} \textbf{76} 012334

\bibitem{deV2} de Vicente J I 2007 \textit{Phys. Rev. A} \textbf{75} 052320

\bibitem{Dat} Datta A, Flammia S T, Shaji A and Caves C M 2007
\textit{Phys. Rev. A} \textbf{75} 062117

\bibitem{mink1} Zhang J, Li C W, Wu J W, Wu R B and Tarn T J 2006 \textit{Phys. Rev. A} \textbf{73}
022319

\bibitem{mink2} Zhang J, Li C W, Tarn T J and Wu J W 2007 \textit{Phys. Rev. A} \textbf{76} 032306

\bibitem{Hor2} Horodecki M, Horodecki P and Horodecki R 2006 \textit{Open Syst. Inf. Dyn.} \textbf{13} 103

\bibitem{Blo} Bloch F 1946 \textit{Phys. Rev.} \textbf{70} 460

\bibitem{Hio} Hioe F T and Eberly J H 1981 \textit{Phys. Rev. Lett.} \textbf{47} 838

\bibitem{Fan} Fano U 1983 \textit{Rev. Mod. Phys.} \textbf{55} 855

\bibitem{PHor} Horodecki P 1997 \textit{Phys. Lett. A} \textbf{232} 333

\bibitem{upb1} Horodecki P, Horodecki M and Horodecki R 1999 \textit{Phys.
Rev. Lett.} \textbf{82} 1056

\nonum Bennett C H, DiVincenzo D P, Mor T, Shor P W, Smolin J A and
Terhal B M 1999 \textit{Phys. Rev. Lett.} \textbf{82} 5385

\nonum Bru{\ss} D and Peres A 2000 \textit{Phys. Rev. A} \textbf{61}
030301(R)

\bibitem{upb2} DiVincenzo D P, Mor T, Shor P W, Smolin J A and
Terhal B M 2003 \textit{Comm. Math. Phys.} \textbf{238} 379

\bibitem{Bre2} Breuer H P 2005 \textit{J. Phys. A: Math. Gen.} \textbf{38}
9019

\bibitem{Aug} Augusiak R and Stasi\'{n}ska J 2007 \textit{Phys. Lett. A}
\textbf{363} 182

\bibitem{normalform} Verstraete F, Dehaene J and De Moor B 2003
\textit{Phys. Rev. A} \textbf{68} 012103

\bibitem{Lei} Leinaas J M, Myrheim J and Ovrum E 2006 \textit{Phys. Rev. A}
\textbf{74} 012313

\bibitem{Ver} Verstraete F 2002 \textit{Ph. D. Thesis} (Katholieke Universiteit
Leuven)

\bibitem{upb3} DiVincenzo D P and Terhal B M 2001 \textit{Proc. of the
XIII Int. Cong. on Mathematical Physics (London 2000)} (Boston: Int.
Press) p 399

\bibitem{EOF} Bennett C H, DiVincenzo D P, Smolin J A and Wootters W K,
\textit{Phys. Rev. A} \textbf{54} 3824

\bibitem{Woo} Wootters W K 1998 \textit{Phys. Rev. Lett.} \textbf{80} 2245

\bibitem{Run1} Rungta P, Bu\v{z}ek V, Caves C M, Hillery M and
Milburn G J 2001 \textit{Phys. Rev. A} \textbf{64} 042315

\bibitem{Run2} Rungta P and Caves C M 2003 \textit{Phys. Rev. A} \textbf{67} 012307

\bibitem{Alb} Albeverio S and Fei S M 2001 \textit{J. Opt. B: Quantum
Semiclass. Opt.} \textbf{3} 223

\bibitem{Osb} Osborne T J 2005 \textit{Phys. Rev. A} \textbf{72} 022309

\bibitem{Kot} Kothe C and Bj\"ork G 2007 \textit{Phys. Rev. A} \textbf{75} 012336


\end{thebibliography}
\end{document}